\documentclass[10pt]{article}
\usepackage{amsmath}
\usepackage{amssymb}
\usepackage{graphicx}
\usepackage{cite}
\usepackage{color}
\usepackage[labelfont=bf,labelsep=period,justification=raggedright]{caption}

\makeatletter
\renewcommand{\@biblabel}[1]{\quad#1.}
\makeatother

\date{}

\begin{document}

\begin{flushleft}
{\Large
\textbf{Win-stay-lose-learn promotes cooperation in the spatial prisoner's dilemma game}
}\sffamily
\\[3mm]
\textbf{Yongkui Liu,$^{1,2,3,\ast}$ Xiaojie Chen,$^{4,\dagger}$ Lin Zhang,$^{1}$ Long Wang,$^{5}$ Matja{\v z} Perc$^{6,\S}$}
\\[2mm]
{\bf 1} School of Automation Science and Electrical Engineering, Beihang University, Beijing, China
{\bf 2} School of Electronic and Control Engineering, Chang'an University, Xi'an, China
{\bf 3} Center for Road Traffic Intelligent Detection and Equipment Engineering, Chang'an University, Xi'an, China
{\bf 4} Evolution and Ecology Program, International Institute for Applied Systems Analysis, Laxenburg, Austria
{\bf 5} State Key Laboratory for Turbulence and Complex Systems, College of Engineering, Peking University, Beijing, China
{\bf 6} Department of Physics, Faculty of Natural Sciences and Mathematics, University of Maribor, Slovenia
\\[2mm]
$^{\ast}$ykliu@chd.edu.cn\\
$^{\dagger}$chenx@iiasa.ac.at\\
$^{\S}$matjaz.perc@uni-mb.si [www.matjazperc.com]
\end{flushleft}
\sffamily
\section*{Abstract}
Holding on to one's strategy is natural and common if the later warrants success and satisfaction. This goes against widespread simulation practices of evolutionary games, where players frequently consider changing their strategy even though their payoffs may be marginally different than those of the other players. Inspired by this observation, we introduce an aspiration-based win-stay-lose-learn strategy updating rule into the spatial prisoner's dilemma game. The rule is simple and intuitive, foreseeing strategy changes only by dissatisfied players, who then attempt to adopt the strategy of one of their nearest neighbors, while the strategies of satisfied players are not subject to change. We find that the proposed win-stay-lose-learn rule promotes the evolution of cooperation, and it does so very robustly and independently of the initial conditions. In fact, we show that even a minute initial fraction of cooperators may be sufficient to eventually secure a highly cooperative final state. In addition to extensive simulation results that support our conclusions, we also present results obtained by means of the pair approximation of the studied game. Our findings continue the success story of related win-stay strategy updating rules, and by doing so reveal new ways of resolving the prisoner's dilemma.

\section*{Introduction}
Evolutionary game theory provides a powerful mathematical framework for studying the emergence and stability of cooperation in social, economic and biological systems \cite{maynard_82, sigmund_93, hofbauer_98, nowak_06, sigmund_10}. The prisoner's dilemma game, in particular, is frequently considered as a paradigm for studying the emergence of cooperation among selfish and unrelated individuals \cite{axelrod_84}. The outcome of the prisoner's dilemma game is governed by pairwise interactions, such that at any instance of the game two individuals, who can either cooperate or defect, play the game against each other by selecting their strategy simultaneously and without knowing what the other player has chosen. Both players receive the reward $R$ upon mutual cooperation, but the punishment $P$ upon mutual defection. If one player defects while the other cooperates, however, the cooperator receives the sucker's payoff $S$ while the defector receives the temptation $T=b$. Since $T>R>P>S$, there is an innate tension between individual interests (the rational strategy, yielding an optimal outcome for the player regardless of what the other player chooses, is defection) and social welfare (for the society as a whole the optimal strategy is cooperation) that may result in the ``tragedy of the commons'' \cite{hardin_g_s68}. Five prominent rules for the successful evolution of cooperation, which may help avert an impeding social decline, are kin selection, direct and indirect reciprocity, network reciprocity as well as group selection, as comprehensively reviewed in \cite{nowak_s06}.

Since the pioneering work of Nowak and May \cite{nowak_n92b} spatial games have received ample attention, and they have become inspirational for generations of researchers trying to reveal new ways by means of which cooperation can prevail over defection \cite{szabo_pr07, roca_plr09, perc_bs10}. In the context of spatial games, network topology and hierarchies have been identified as a crucial determinant for the success of cooperative behavior \cite{santos_prl05, vukov_pre05, santos_pnas06, santos_prsb06, gomez-gardenes_prl07, poncela_njp07, lozano_ploso08, poncela_ploso08, poncela_epl09, poncela_njp09, pacheco_ploscb09, grujic_pone10, dai_ql_njp10, poncela_pre11, lee_s_prl11, dai_ql_pre11}, where in particular the scale-free topology has proven very beneficial for the evolution of cooperation. In fact, payoff normalization \cite{tomassini_ijmpc07, wu_zx_pa07, szolnoki_pa08} and conformity \cite{pena_pre09} belong to the select and very small class of mechanisms that can upset the success of cooperators on such highly heterogeneous networks. Other approaches facilitating the evolution of cooperation include the introduction of noise to payoffs and updating rules \cite{szabo_pre98, szabo_pre05, perc_njp06a, vukov_pre06, szolnoki_pre09d, szabo_epl09}, asymmetry between interaction and replacement graphs \cite{ohtsuki_prl07, ohtsuki_jtb07b}, diversity \cite{hu_pa07, perc_pre08, santos_n08, wu_zx_pre09b}, differences between time scales of game dynamics \cite{roca_prl06, szolnoki_epl08, wu_zx_pre09b, rong_pre10}, as well as adoption of simultaneous different strategies against different opponents \cite{wardil_epl09}. Somewhat more personally-inspired features supporting the evolution of cooperation involve memory effects \cite{liu_yk_pa10}, heterogeneous teaching activity \cite{szolnoki_epl07, szolnoki_njp08, szabo_pre09}, preferential learning \cite{guan_epl06, wu_zx_pre06}, mobility \cite{vainstein_jtb07, helbing_pnas09, meloni_pre09, droz_epjb09, wu_zx_pre09}, myopically selective interactions \cite{chen_xj_pre09}, and coevolutionary partner choice \cite{fu_pre08b, fu_pre09, van-segbroeck_prl09}, to name but a few examples studied in recent years.

Regardless of the details of mechanisms that may promote the evolution of cooperation in the spatial prisoner's dilemma game, most frequently, it was assumed that individual players learn from their neighbors to update their strategies, and that they do so more or less in every round of game. But in reality, we are much less prone to changing our strategy (see \cite{cho_jet05} and references therein). Withstanding ample trial and error, only when we feel to a sufficiently high degree unsuccessful and dissatisfied may we be tempted into altering it. Enter aspirations, which play a pivotal role in determining our satisfaction and the notion of personal success. Indeed, the subtle role of aspirations in evolutionary games has recently received a lot of attention \cite{macy_pnas02, chen_xj_pre08, platkowski_pre09, platkowski_aml09, yang_hx_pre10, zhang_hf_pa10b, wang_z_pre10, perc_pone11, liu_yk_epl11}, and amongst others, it was discovered that too high aspirations may act detrimental on the evolution of cooperation. This result is quite intuitive, as very high aspirations will inevitably lead players to choose defection over cooperation in order to achieve their high goals. Regardless of the level, however, aspiration provides an elegant means to define when a player is prone to changing its strategy and when not. In particular, if the performance trails behind the aspiration, then the player will likely attempt to change its strategy. On the other hand, if the performance agrees or is even better than the aspiration, then the player will not attempt to change its strategy. Here we make use of this simple and intuitive rationale to build on the seminal works that introduced and studied the win-stay-like (win-stay-lose-shift being the most prominent example) strategy updating rules \cite{nowak_n93, posch_jtb99, imhof_jtb07}.

In this paper, we thus introduce a so-called win-stay-lose-learn strategy updating rule as follows. When satisfied, players maintain their strategies and do not attempt to change them. When dissatisfied, however, players proceed as it is traditionally assumed, i.e., by attempting to imitate the strategy of one of their neighbors \cite{matsen_pnas04}. It is worth pointing out that since in evolutionary games on structured populations individuals need to interact with their neighbors for collecting payoffs, it can be assumed that such an interaction mode provides enough opportunities for players to observe the information of their neighbors (including payoffs). Nevertheless, it is difficult to pinpoint whether our model can be held accountable only for human behavior or also for animal behavior. Certainly, some level of intelligence is needed by the players to accommodate our assumptions. In the proposed model the strategy updating is thus conditional on whether the players are satisfied or not, which we determine by means of the aspiration level $A>0$ that is considered as a free parameter. We note again that the majority of previous studies assumed that players will always try and adopt the strategy of one of their neighbors, even if the neighbor is performing worse, and regardless of the individual level of satisfaction. Here we depart from this somewhat simplifying assumption, and by doing so discover that the aspiration-based conditional strategy updating, termed win-stay-lose-learn, strongly promotes the evolution of cooperation in the spatial prisoner's dilemma game, even under very unfavorable initial conditions \cite{langer_jtb08, chen_xj_pla08} and by high temptations to defect $b$. All the details of the proposed win-stay-lose-learn strategy updating rule and the setup of the spatial prisoner's dilemma game are described in the Methods section, while here we proceed with presenting the main results.

\section*{Results}
\begin{figure}
\begin{center}\includegraphics[width=14cm]{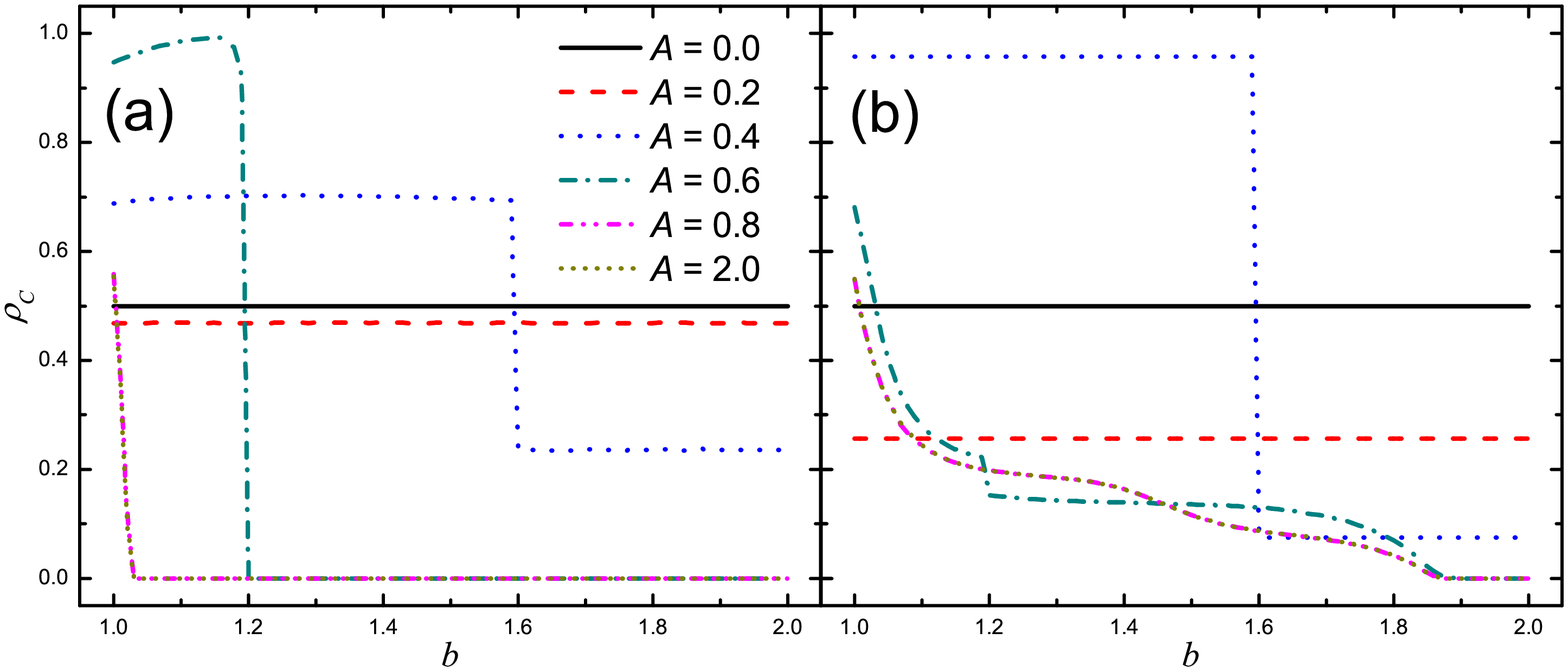}\end{center}
\caption{\textbf{Win-stay-lose-learn promotes the evolution of cooperation, especially if intermediate aspirations determine the satisfiability of players.} Presented is the stationary fraction of cooperators $\rho_C$ in dependence on the temptation to defect $b$ for different values of the aspiration $A$, as obtained by means of simulations (panel a) and the pair approximation (panel b). By comparing the results presented in the two panels, it can be observed that the pair approximation is to a large degree successful in reproducing the qualitative features of the simulations.}
\label{fig1}
\end{figure}

We start by presenting the results as obtained when the cooperators and defectors are distributed uniformly at random, each thus initially occupying half of the square lattice. As the main parameters, we consider the aspiration level $A$ and the temptation to defect $b$. Figure~\ref{fig1} shows the fraction of cooperators $\rho_C$ as a function of the temptation to defect $b$ for different aspiration levels $A$. We find that the aspiration level has a significant influence on the density of cooperators. In particular, for small values of $A$, intermediate levels of cooperation are maintained, and the temptation to defect has no effect on cooperation, e.g., for $A=0$ and $A=0.2$, $\rho_C=0.5$ and $\rho_C=0.47$, respectively, irrespective of the value of $b$. When $A$ is within an intermediate range, the density of cooperators increases with increasing $A$, however, the maximal value of $b$ still warranting high cooperation levels becomes smaller, e.g., when $A=0.4$, $\rho_C=0.7$ for $b\in [0,1.6]$ and when $A=0.6$, $\rho_C\rightarrow 1.0$ for $b\in [0,1.2]$. In addition, as $b$ increases, transitions to different stationary states can be observed for certain values of $A$. Notable examples occur at $b=1.6$ for $A=0.4$ and at $b=1.2$ for $A=0.6$, which will be explained below. When $A=0.8$, cooperation cannot evolve even if the value of $b$ is only slightly larger than $1.0$. When $A=2.0$, the result coincides with that for $A=0.8$. It is worth pointing out that, in fact, when $A$ is large, e.g., $A=2.0$, individuals are always dissatisfied (as shown below), and that then our model recovers the traditional version of the prisoner's dilemma game. By comparing the results for $A=2.0$ and those for other values of $A$, as shown in Fig.~\ref{fig1}, we find that the present updating rule can effectively facilitate the evolution of cooperation. In particular, when $A\geq2.0$, cooperators can survive only if $b<1.05$. By contrast, with win-stay-lose-learn updating, cooperators can not only survive but also thrive even for much larger values of $b$.

The results obtained by means of pair approximation are presented in Fig.~\ref{fig1}(b). It can be observed that the pair approximation can qualitatively correctly predict the cooperation level, especially for small values of $A$. For example, the result for $A=0.0$ is exactly the same as the simulation result. The transition at $A=0.4$ can be observed in Fig.~\ref{fig1}(b). On the other hand, when $A>0.5$, the satisfaction of individuals is increasingly difficult to achieve such that individuals tend to learn their neighbors when updating strategies, and the current model approaches the model of continuous updating \cite{guan_epl06}. Hence, there exist some differences between the results obtained by means of simulations and the pair approximation. Despite this, in general the results of the pair approximation support the main conclusions at which we arrive at by means of simulations.

\begin{figure}
\begin{center}\includegraphics[width=8cm]{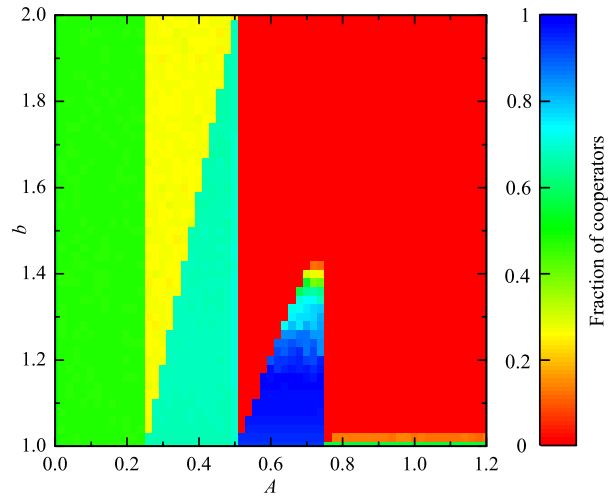}\end{center}
\caption{\textbf{Transitions from predominantly defective to predominantly cooperative states in dependence on the aspiration level $A$ and the temptation to defect $b$.} Presented is the color-coded (see bar on the right) fraction of cooperators. The multitude of transitions in the color map points towards a high complexity of the underlying mechanisms warranting highly cooperative states (see main text for details).}
\label{fig2}
\end{figure}

In order to obtain a more complete picture about the joint effects of the aspiration level and the temptation to defect, we show the simulation results as a function of both $A$ and $b$, as shown in Fig. \ref{fig2}. The results are consistent with those presented in Fig.~\ref{fig1}(a), e.g., when $A\leq0.25$, an intermediate level of cooperation ($\rho_C\approx0.5$) is maintained, irrespective of the value of $b$. Within the interval of $0.25<A<0.5$, the cooperation level is higher than that for $A\leq0.25$ and the transition can be observed at a fixed value of $b$ for each value of $A$. It is interesting that the highest levels of cooperation occur within the interval of $0.5\leq A<0.75$. Moreover, it can be observed that as $A$ increases, discontinuous transitions occur at $A=0.0$, $0.25$, $0.5$ and $0.75$.

These transitions can be understood as follows. On a square lattice with nearest neighbor interactions, the payoffs of a cooperator and a defector are given by $n_1R+n_2S$ and $n_3T+n_4P$, respectively, where $n_k\in \{0,1,2,3,4\}$, and $k\in \{1,2,3,4\}$. Given that $T=b$, $R=1$, and $P=S=0$, the above payoffs can be simplified as $n_1$ and $n_3b$, respectively. In our model, when an individual is dissatisfied, it will learn a randomly chosen neighbor, which may lead to the change of $\rho_C$. For a cooperator, when $n_1<4A$, it is dissatisfied. While for a defector, the condition for its dissatisfaction is $n_3b<4A$. The phase transition points can be obtained by letting $n_1=4A$ and $n_3b=4A$. Thus, the value of $A$ at which phase transition occurs is given by $A=n_1/4$ and that for $b$ is given by $b=(4A)/n_3>1$. Considering all the possible values of $n_1$, that is $n_1=0$, $1$, $2$, $3$ and $4$, we can obtain the phase transition points of $A$, which are $A=0.0$, $0.25$, $0.5$, $0.75$ and $1.0$, as shown in Fig.~\ref{fig2}. (As a matter of fact, $A=1.0$ is also a phase transition point, however, because when $A=1.0$, the density of cooperators is very low such that the phase transition phenomenon cannot be observed). The phase transition points of $b$ can be calculated similarly, e.g., when $n_3=1$ and $A=0.4$, the phase transition point of $b$ is $b=1.6$, which is verified by our simulation (see Fig.~\ref{fig2}).

\begin{figure}
\begin{center}\includegraphics[width=8cm]{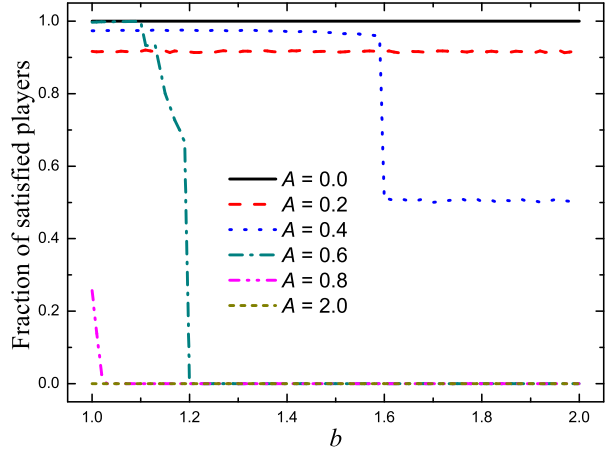}\end{center}
\caption{\textbf{The fraction of satisfied players decreases with increasing aspirations.} Presented is the fraction of satisfied players in the population, for which it holds that $P_i \geq A_i$, in dependence on the temptation to defect $b$ for different values of the aspiration $A$. It is interesting to observe that for low values of $A$ the fraction of satisfied players is independent of $b$, while for intermediate and large values of $A$ it decreases with increasing $b$. Also note that for $A=0.0$ ($A=2.0$) all (no) players are satisfied.}
\label{fig3}
\end{figure}

Since the strategy changes of individual players are determined by their satisfaction, we proceed with the results on the satisfaction rates in the population as a function of $b$ for different values of $A$, as shown in Fig.~\ref{fig3}. A highly cooperative society where each member is satisfied can be declared as the ultimate goal. If all members cooperate, then the social welfare will peak. Moreover, if then every member is satisfied, the society will be stable. We find from Fig.~\ref{fig3} that, if we regard the present system as a social prototype, then the optimal situation occurs within the interval of $b\in[1.0,1.2]$ for $A=0.6$, since it leads to a highly cooperative society with a high satisfaction rate. For the extreme case of $A=0.0$, all individuals are satisfied. On the contrary, at the other extreme, i.e.,at $A=2.0$, no individual is ever satisfied. For $A=0.2$, even though that more than $90\%$ of individuals are satisfied, the cooperation level is not high ($\rho_C=0.47$). This indicates that a large number of defectors are satisfied by exploiting cooperators. The obtained result for $A=0.2$ reveals that a society where each member has a low aspiration level cannot be cooperative due to an inherent lack of incentives. When $A=0.4$, the fraction of satisfied individuals drops suddenly at $b=1.6$. When $b\leq1.6$, nearly $70\%$ of individuals cooperate and the satisfaction rate is high, which is more or less a better situation. Whereas $b>1.6$ results in the low cooperation level as well as the low satisfaction rate, which is a society that should be avoided. When $A=0.8$, few individuals in the population are satisfied such that almost all individuals seek for higher payoffs by imitation, and ultimately defection becomes the unique choice. This confirms the standard game theoretical result that, in a society where individuals imitate each other, individual greediness (characterized by too high aspiration) may hinder the emergence of cooperation and eventually harm the benefit of each member of the society.

\begin{figure}
\begin{center}\includegraphics[width=14cm]{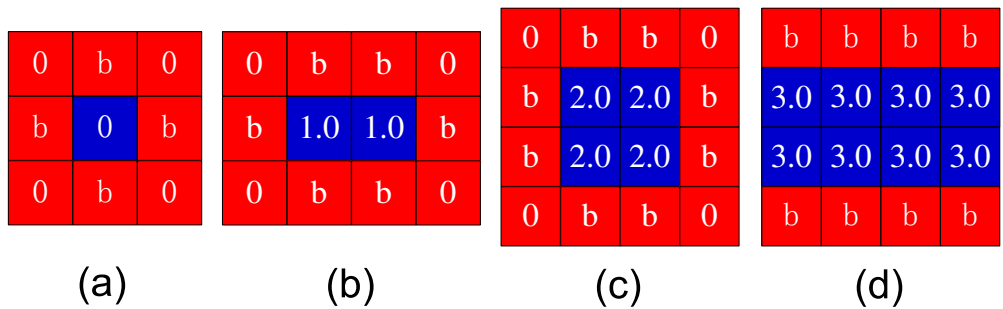}\end{center}
\caption{\textbf{Special initial configurations of cooperators reveal their potential to expand into the territory of defectors.} In all panels the cooperators are depicted blue while defectors are depicted red. Each small square corresponds to a single player. Denoted values correspond to the payoffs of individual players, as obtained for the presented configurations. See also Fig.~\ref{fig5} for related results.}
\label{fig4}
\end{figure}

\begin{figure}
\begin{center}\includegraphics[width=13.5cm]{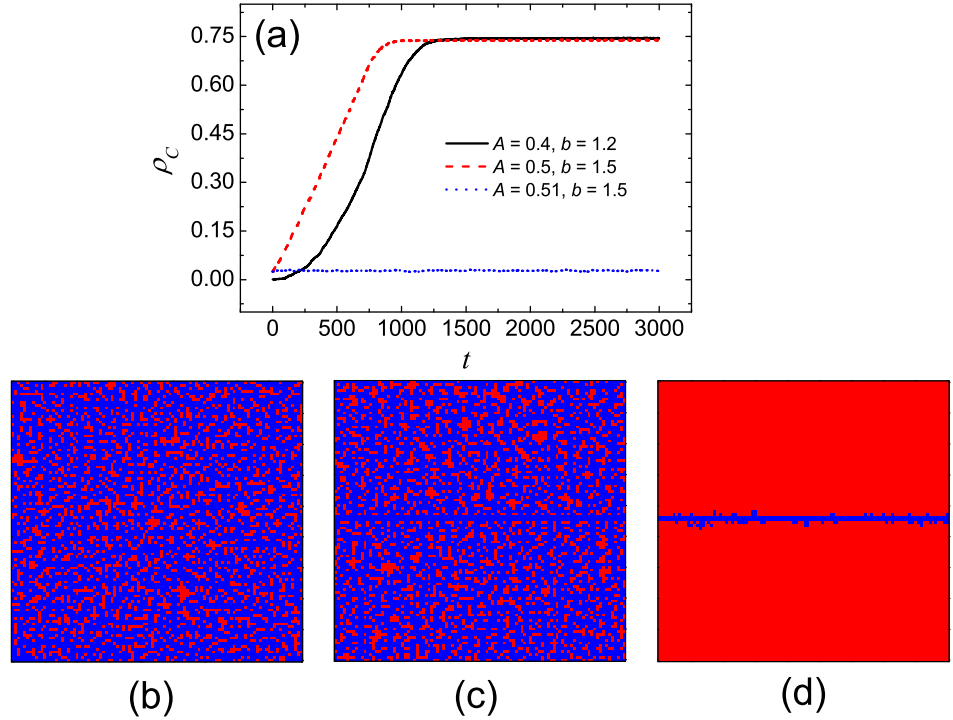}\end{center}
\caption{\textbf{Robustness of the evolution of cooperation under adverse initial conditions.} Panel (a) features the time evolution of the fraction of cooperators for different combinations of $A$ and $b$, as obtained when using the initial conditions presented in Figs.~\ref{fig4}(c) (black solid line) and (d) (dashed red and dotted blue line). Bottom row features the characteristic snapshots of the spatial grid (cooperators are blue, defectors are red), corresponding to the black solid line (panel b), the dashed red line (panel c), and the dotted blue line (panel d). It can be observed that cooperators may significantly outnumber defectors in the stationary state, even if starting from highly unfavorable conditions, as long as the aspirations are appropriately adjusted.}
\label{fig5}
\end{figure}

Because initial conditions are relevant for the evolutionary success of cooperators in spatial games \cite{langer_jtb08, chen_xj_pla08}, it is also of interest to test the robustness of the proposed updating rule. We thus investigate how cooperation evolves under different (adverse) initial conditions, which are shown in Fig.~\ref{fig4}. We first focus on the initial configuration of cooperators and consider the case of Fig.~\ref{fig4}(a), where only one cooperator exists in the population initially. For $A>0$, the cooperator surely resorts to defection by imitation because of his dissatisfaction. Hence, a single cooperator surrounded by defectors cannot survive $A>0$. When initially there are two neighboring cooperators [see Fig.~\ref{fig4}(b)], for $A\leq0.25$, both cooperators and defectors at the boundary are satisfied such that the pattern is stable. However, when $0.25< A \leq b/4$, defectors are satisfied but cooperators are not. Thus, cooperators will become defectors by imitation. When $A>b/4$, all individuals are dissatisfied such that all of them imitate neighbors' strategies. Since defectors have a higher payoff, cooperators are more likely to become defectors. Therefore, $A\leq0.25$ is needed to make cooperators survive. When there exist four cooperators in the population initially, as shown in Fig.~\ref{fig4}(c), the pattern is frozen if $A\leq b/4$. However, when $b/4<A\leq 0.5$, cooperators can have advantages over defectors such that they can invade defectors and dominate the population ultimately, as shown in Fig.~\ref{fig5}. This indicates the relevance of our model since it allows cooperators to thrive in harsh conditions where there only exist several cooperators initially. When $A>0.5$, cooperators cannot expand their territories and, at the same time, they confront the intense invasion by defectors. Eventually cooperators are wiped out from the population. In this scenario, $A\leq 0.5$ is needed to maintain the pattern. This indicates that greediness may be detrimental to the emergence of cooperation. The more favorable case emerges when each cooperator has three cooperative neighbors, \emph{i.e.} the population is initialized with two neighboring straight lines of cooperators, as shown in Fig.~\ref{fig4}(d). Under these circumstances, $A\leq 0.75$ is required for cooperators to maintain strategies, which, however, does not warrant cooperators to expand territories. In order to realize the expansion, we need $A\leq 0.5$. When $A>0.5$, the expansion of areas of cooperators is significantly restrained, which is demonstrated in Fig.~\ref{fig5}. One can find that in order for the boom of cooperation, cooperators must, first, form clusters, which ensures they have advantages over defectors in terms of payoffs. Second, cooperators cannot set their aspirations too high such that their satisfactions can easily be achieved \cite{roca_pnas11}. Thus, they will hold their strategies, which is the precondition for the spreading of strategies. The fulfillment of the above two conditions as well as the dissatisfaction of defectors leads to the dissemination of the cooperative strategy in the population.

\begin{figure}
\begin{center}\includegraphics[width=13.2cm]{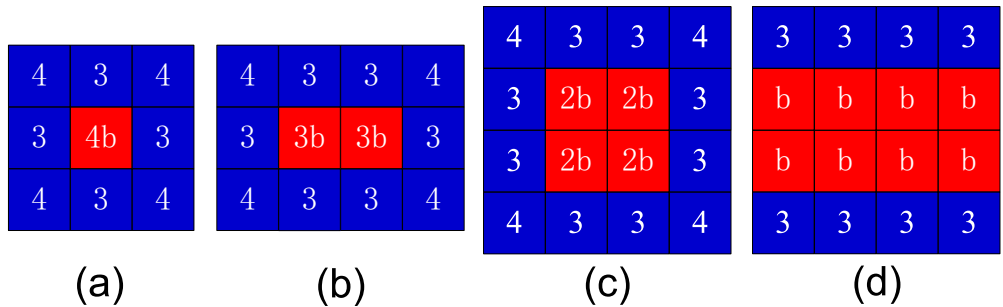}\end{center}
\caption{\textbf{Special initial configurations of defectors reveal their potential to invade cooperators.} In all panels the cooperators are depicted blue while defectors are depicted red. Each small square corresponds to a single player. Denoted values correspond to the payoffs of individual players, as obtained for the presented configurations (see main text for details).}
\label{fig6}
\end{figure}

Lastly, we elaborate on how cooperators can resist the invasion by defectors. For this purpose, we consider the special initial conditions depicted in Fig.~\ref{fig6}. Focusing first on Fig.~\ref{fig6}(a), we find that when $A>0.75$, cooperators are dissatisfied, which can lead to the extinction of cooperators. The case in Fig.~\ref{fig6}(b) is qualitatively the same, the only difference being that the payoff of each defector is $3b$. In the situation where there exists a square block of four defectors [Fig.~\ref{fig6}(c)], when $A\leq 0.75$, cooperators can at least survive. If $b/2<A\leq 0.75$ ($b<1.5$), cooperators can even invade the dissatisfied defectors. If $b>1.5$, cooperators and defectors can coexist. On the contrary, when $A>0.75$, cooperators are doomed to extinction. The above analysis explains why when $A>0.75$, cooperators cannot flourish. That is, as long as cooperators do not set their goals too high (too greedy), cooperators can resist the invasion of defectors \cite{roca_pnas11}. If they can also have higher payoffs than neighboring defectors, as shown in Figs.~\ref{fig6}(c) and (d), then defectors can be defeated.

\section*{Discussion}
In summary, we have studied the impact of the win-stay-lose-learn strategy updating rule on the evolution of cooperation in the spatial prisoner's dilemma game. Unlike in the majority of previous works, in our case the strategy updating is not unconditional, but rather it depends on the level of satisfaction of individual players. The latter is determined by the aspiration level, which we have considered as a free parameter. If the payoff of a player is equal or higher than its aspiration, it is assumed that this player is satisfied and that there is thus no immediate need of changing its strategy. Conversely, if the payoff is lower than the aspired amount, the player will attempt to adopt the strategy of one of its nearest neighbors in the hope that it will reach the desired success. With this setup, we have found that if all players retain their strategies when being satisfied then the evolution of cooperation is remarkably facilitated. Especially for intermediate values of the aspiration parameter, e.g., $A=0.6$, virtually complete cooperation dominance can be achieved even for values of the temptation to defect that significantly exceed $1$. This is in sharp contrast to the results obtained with (too) large aspiration levels, e.g., $A=2.0$, where the traditional version of the spatial prisoner's dilemma game is essentially fully recovered. The presented results also indicate that as long as individuals are not too greedy, i.e., aspire to modest (honest) incomes, cooperation thrives best, which is also in agreement with recent results obtained by means of a different model \cite{roca_pnas11}. Moreover, we have tested the impact of different initial configurations, in particular such where cooperators initially have an inherent disadvantage over defectors, and we have discovered that the studied win-stay-lose-learn rule ensures that cooperators are able to spread even from very small numbers. In this sense, the proposed rule is very effective in unleashing the spreading potential of cooperative behavior, which is to some extent already provided (seeded) by means of spatial reciprocity \cite{nowak_n92b}. We have also employed the pair approximation method to support our simulation results with semi-analytical calculations and to explain the observed transitions to different levels of cooperation on the square lattice.

It is instructive to discuss the differences between this work and related previous works \cite{chen_xj_pre08,platkowski_pre09,zhang_hf_pa10b,wang_z_pre10,perc_pone11,liu_yk_epl11}. For example, Chen and Wang \cite{chen_xj_pre08} investigated a stochastic win-stay-lose-shift (WSLS) rule, under which dissatisfied individuals switch their strategies to the opposite one. It was reported that for small values of the temptation to defect cooperation can be best promoted at intermediate values of the aspiration level. Moreover, in \cite{platkowski_pre09} a N-person prisoner's dilemma game in a continuous population with a time-dependent aspiration level was investigated, while in \cite{zhang_hf_pa10b} each individual had an aspiration-based learning motivation (which actually can depend directly on the aspiration level according to the rule of WSLS). It was reported that the results produced in \cite{zhang_hf_pa10b} are similar to those in \cite{chen_xj_pre08}. In \cite{perc_pone11} a payoff-based preferential learning mechanism was investigated (where individuals with higher payoffs are more likely to be imitated), and in \cite{liu_yk_epl11} an aspiration-based preferential learning mechanism was studied where an individual whose strategy can provide the desired payoff when being imitated will be imitated also in the next round. In our model, however, we incorporate individual aspirations into the traditional imitation rule, and investigate how cooperation evolves under the aspiration-based conditional learning in the spatial prisoner's dilemma game. The proposed rule is simple and reasonable, and moreover, we show that it is effective and robust in promoting cooperation. In particular, cooperation can be maintained (and can even thrive) even under unfavorable initial conditions.

Our work also has some parallels with other models not necessarily incorporating aspirations. Namely, the model introduced recently in \cite{liu_rr_epl10}, where inertia was considered as something that can disable players to actively change their strategies, or with the early win-stay-lose-shift models \cite{nowak_n93, posch_jtb99, imhof_jtb07}. Furthermore, it is possible to relate our work to those considering the importance of time scales in evolutionary games \cite{roca_prl06, wu_zx_pre09b, rong_pre10}. Note that under the presently introduced win-stay-lose-learn rule, for small values of $A$, the strategy transfers are rare and far apart in time. This has similar consequences as when decreasing the strategy-selection rate \cite{rong_pre10}. For intermediate aspiration levels, however, we essentially have a segregated population if judging from individual satisfaction, i.e., some players are satisfied while others are not. This in turn implies that different players have different strategy-selection time scales. Even more importantly, these time scales are adaptive (they change over time), as players that eventually do change their strategies may go from being dissatisfied to becoming satisfied, or vice versa. In this sense, our model introduces different evolutionary time scales by means of aspiration, which is an endogenous property of individuals, and in so doing, it also relaxes the demand for their (the players') rationality. Note that in our model, there is frequently no need to compare payoffs with the neighbors, apart from when approaching the $A \to 2$ limit. Lastly, we hope that this study will enrich our knowledge on how to successfully resolve the prisoner's dilemma, and we hope it will inspire further work along this very interesting and vibrant avenue of research.

\section*{Methods}
The spatial prisoner's dilemma game is staged on a square lattice of size $L \times L$ with periodic boundary conditions. In accordance with common practice \cite{nowak_n92b}, the payoffs are as follows: $T=b$ is the temptation to defect, $R=1$ is the reward for mutual cooperation, while $P=S=0$ are the punishment for mutual defection and the sucker's payoff, respectively, where $1 < b < 2$. Although this formulation of the game has $P=S$ rather than $P>S$, it captures succinctly the essential social dilemma, and accordingly, the presented results can be considered fully relevant and without loss of generality with respect to more elaborated formulations of the payoffs. Moreover, each player $i$ has an aspiration level $A_{i}=k_i A$, where $k_i$ is the player's degree and $A$ is a free parameter that determines the overall aspiration level of the population, which is typically constrained to the interval $0<A<b$. Since we consider the square lattice as the interaction network, we have $k_i=k=4$, which in turn postulates that each player in this study has an equal aspiration equal to $k A$.

Player $i$ acquires its payoff $P_i$ by playing the game with its four nearest neighbors. A randomly selected nearest neighbor $j$ acquires its payoff $P_j$ likewise by playing the game with its four nearest neighbors. If $P_i \geq A_i$, i.e., if the payoff of player $i$ is equal or higher than its aspiration, then strategy adoption from player $j$ is not attempted. If, however, $P_i < A_i$, then player $i$ adopts the strategy of player $j$ with the probability
\begin{equation}
W=\frac{1}{1+\exp[(P_i-P_j)/\kappa]}
\end{equation}
where $\kappa$ determines the amplitude of noise \cite{szabo_pre98}, accounting for imperfect information and errors in decision making. It is well-known that there exists an optimal intermediate value of $\kappa$ at which the evolution of cooperation is most successful \cite{szabo_pre05, perc_njp06a}, yet in general the outcome of the prisoner's dilemma game is robust to variations of $\kappa$. Without much loss of generality, we use $\kappa=0.1$, meaning that it is very likely that the better performing players will pass their strategy to other players, yet it is not impossible that players will occasionally learn also from the less successful neighbors. The simulations of this spatial prisoner's dilemma game were performed by means of a synchronous updating rule, using $L=100$ to $400$ system size and discarding the transient times prior to reaching the stationary states before recording the average fraction of cooperators $\rho_C$ in the population. We have verified that the presented results are robust to variations of the system size, and to the variation of the simulation procedure (e.g., by using random rather than synchronous updating). It is also worth noting that because $A<b$ and $b<2$, the present definition of the win-stay-lose-learn transforms to the traditional spatial prisoner's dilemma game when $A \geq 2.0$, given that then individual cannot be satisfied and thus attempt to change their strategy whenever they receive a chance to do so.

In addition to the simulation results of the proposed spatial game, we also present the results of pair approximation \cite{hauert_n04, hauert_ajp05, guan_epl06, wu_zx_pre07, chen_xj_pre08b} that are obtained with the rate equations of cooperator-cooperator ($c,c$) and cooperator-defector ($c,d$) edges, which are as follows:
\begin{eqnarray}
\dot{p}_{c,c}&=&\sum_{x,y,z}[n_c(x,y,z)+1]p_{d,x}p_{d,y}p_{d,z}\times \nonumber\\
&&\sum_{u,v,w}p_{c,u}p_{c,v}p_{c,w} f[P_d(x,y,z),P_c(u,v,w)] \nonumber\\
&& -\sum_{x,y,z}n_c(x,y,z)p_{c,x}p_{c,y}p_{c,z}\times  \nonumber\\
&& \sum_{u,v,w}p_{d,u}p_{d,v}p_{d,w} f[P_c(x,y,z),P_d(u,v,w)],
\end{eqnarray}
\begin{eqnarray}
\dot{p}_{c,d}&=&\sum_{x,y,z}[1-n_c(x,y,z)]p_{d,x}p_{d,y}p_{d,z}\times \nonumber\\
&&\sum_{u,v,w}p_{c,u}p_{c,v}p_{c,w} f[P_d(x,y,z),P_c(u,v,w)] \nonumber\\
&& -\sum_{x,y,z}[2-n_c(x,y,z)]p_{c,x}p_{c,y}p_{c,z}\times  \nonumber\\
&& \sum_{u,v,w}p_{d,u}p_{d,v}p_{d,w} f[P_c(x,y,z),P_d(u,v,w)],
\end{eqnarray}
where $x,y,z$ are either cooperators or defectors and $n_c(x,y,z)$ denote the number of cooperators among $x,y,z$. Moreover,
\begin{eqnarray}
f(P_i,P_j)=
\begin{cases}
  \frac{1}{1+\textmd{exp}[(P_i-P_j)/\kappa]}, \ \ & P_i<A_{i}\\
¡¡0,  \ \ & P_i\geq A_{i}
\end{cases}
\end{eqnarray}
where $P_i$ and $P_j$ are the payoffs of the two neighboring players $i$ and $j$, respectively, and $A_{i}$ is the payoff aspiration of player $i$ (equal to $A k$ for all $i$). By performing the numerical integration for the above two differential equations (2,3), and by using $p_{c,d}=p_{d,c}$ and $p_{c,c}+p_{c,d}+p_{d,c}+p_{d,d}=1$, we can obtain $\rho_C$ from $p_{c,c}+p_{c,d}$.

\clearpage

\end{document}